\documentclass[conference]{IEEEtran}
\IEEEoverridecommandlockouts

\usepackage{subfigure}
\usepackage{adjustbox}
\usepackage{multirow}
\def\BibTeX{{\rm B\kern-.05em{\sc i\kern-.025em b}\kern-.08em
    T\kern-.1667em\lower.7ex\hbox{E}\kern-.125emX}}

\begin{document}

\title{Can gender categorization influence the perception of animated virtual humans?}

\author{\IEEEauthorblockN{Victor Flávio de Andrade Araujo}
\IEEEauthorblockA{\textit{School of Technology} \\
\textit{Pontifical Catholic University} \\
\textit{of Rio Grande do Sul}
\\
\textit{VHLab} \\
Porto Alegre \\ Brazil
\\ victor.flavio@acad.pucrs.br}

\\

\IEEEauthorblockN{Angelo Brandelli Costa}
\IEEEauthorblockA{\textit{School of Health and Life Sciences} \\
\textit{Pontifical Catholic University} \\
\textit{of Rio Grande do Sul}
\\
\textit{PVPP} \\
Porto Alegre \\ Brazil
\\  angelo.costa@pucrs.br}
\and

\IEEEauthorblockN{Diogo Hartmann Muller Schaffer}
\IEEEauthorblockA{\textit{School of Technology} \\
\textit{Pontifical Catholic University} \\
\textit{of Rio Grande do Sul}
\\
\textit{VHLab} \\
Porto Alegre \\ Brazil
\\  diogo.schaffer@acad.pucrs.br}
\\

\IEEEauthorblockN{Soraia Raupp Musse}
\IEEEauthorblockA{\textit{School of Technology} \\
\textit{Pontifical Catholic University} \\
\textit{of Rio Grande do Sul}
\\
\textit{VHLab} \\
Porto Alegre \\ Brazil
\\  soraia.musse@pucrs.br}
}

\maketitle

\begin{abstract}

Animations have become increasingly realistic with the evolution of Computer Graphics (CG). In particular, human models and behaviors were represented through animated virtual humans, sometimes with a high level of realism. In particular, gender is a characteristic that is related to human identification, so that virtual humans assigned to a specific gender have, in general, stereotyped representations through movements, clothes, hair and colors, in order to be understood by users as desired by designers. An important area of study is finding out whether participants' perceptions change depending on how a virtual human is visually presented. Findings in this area can help the industry to guide the modeling and animation of virtual humans to deliver the expected impact to the audience.
In this paper, we reproduce, through CG, a perceptual study that aims to assess gender bias in relation to a simulated baby. In the original study, two groups of people watched the same video of a baby reacting to the same stimuli, but one group was told the baby was female and the other group was told the same baby was male, producing different perceptions. The results of our study with virtual babies were similar to the findings with real babies. First, it shows that people's emotional response change depending on the character gender attribute, in this case the only difference was the baby's name. 
Our research indicates that by just informing the name of a virtual human can be enough to create a gender perception that impact the participant emotional answer.

\end{abstract}



\section{Introduction}
\label{sec:introduction}

Computer Graphics (CG) has evolved in recent years. This evolution helped in the development of more realistic CG characters in terms of their looks and behaviors, bordering on a high level of human likeness and making the audience feel more comfortable with them~\cite{katsyri2015review}.
According to the Uncanny Valley (UV) theory~\cite{mori2012uncanny},
the more realistic an artificial human is, the more likely it is to cause discomfort to the observer. According to~\cite{katsyri2015review}, this feeling of strangeness 
is related to the perception of human likeness
, being an identification issue, i.e., the humans perceive realistic characteristics in the virtual humans. Following this line
, one of these characteristics is gender. According to~\cite{scott2007gender, wallach2010gender}, the term gender is related to social and cultural constructions 
about "appropriate roles" for women and men.

\cite{draude2011intermediaries} claimed that simulated human likeness (in terms of visuals and behaviors) can utilize human self-reliability
, turning something abstract into comfortable. With that, Draude made a provocation, which raised the hypothesis that most artificial beings 
can be assigned a female gender to ensure that the audience feels comfortable. In this sense, in the work of~\cite{araujo2021analysis}, 
results showed that women can feel more comfortable with realistic female characters than realistic male characters, while men feel comfortable similarly about realistic female and male characters. 
Furthermore, following the concept of charismatic leadership~\cite{weber1947legitimate, spencer1970weber, west1980charisma} and relating it to the charisma presented by characters created using CG from different media, the results of Araujo et al. showed that both women and men perceived realistic female characters as more charismatic than realistic male characters. In this case, 
the characters were visibly male and female, but would these results be similar if the virtual human had no gender identification? 
Still in relation to gender categorization,~\cite{draude2011intermediaries} made an analogy, stating that Computer Science has the potential to deconstruct the gender binary. 
What about virtual humans? Is there this potential for gender deconstruction in a genderless virtual human?

According to~\cite{condry1976sex}, children often look for an answer about how they should behave by looking at the ways adults react to them. 
Hypothetically, the chances of observing stereotypes in young children are lower. 
The study with babies is relevant because they are accepted as naturally ''genderless'' in terms of their face and body models, so the bias measure can be considered with less interference.
There are several games in which the player can choose the design of a character
\footnote{https://twinfinite.net/2018/08/games-nailed-character-creation-systems/}\footnote{https://starloopstudios.com/character-design-in-video-games-types-classes-and-characteristics/}. 
In that games, users can create and customize their avatars, however results showed that although some users try to model a 
no defined gender, 
the created type was mostly a male character, according to human perception. The preference to categorize a non-gendered character into a male character can be a problem of gender bias~\cite{nass2000machines}. 
In the movies industry, specifically when we talk about cartoons, female characters are usually more stereotyped~\cite{garza2019emotional} both in terms of visuals\footnote{https://thedissolve.com/news/655-disney-animator-says-women-are-hard-to-draw-becaus/} and movement. 
But are these exaggerations really necessary for the recognition of a female character?
To answer some of these questions,
this work replicates and expands the work of~\cite{condry1976sex}, in which the authors measured the perception of men and women about a video containing a real baby reacting to different stimuli. The authors separated the participants into two groups, one receiving the video of the baby with a male name, and the other receiving the baby with a female name. 
The results showed that women 
perceived more emotions in the baby with a female name, while men 
perceived more emotions in the baby with a male name. The research also showed 
that people's perception is influenced by the gender characteristics, even if that characteristic is just the categorization of the gender or the name. 
We replicate Condry and Condry's experiment using a virtual baby, and also expand the experiment using a third group of people who receives the unnamed baby, in addition to the other two groups that receive the baby with a female and male names.
Therefore, to help answer our questions, the following hypotheses are created and are discussed in this text.

\begin{itemize}
\item $H0_{1}$ defining that women and men perceive emotions similarly in female, male and unnamed animated virtual babies;
\item $H0_{2}$ defining that the unnamed baby is recognized as genderless; and
\item $H0_{3}$ defining that women and men perceive comfort and charisma similarly in baby with male name and unnamed baby, and different in relation to baby with female name, (following the results of work of~\cite{araujo2021analysis}).
\end{itemize}

The main contributions of this paper are: the investigation of human bias attributing gender to the virtual humans, and its impact on the participants perceived emotions, comfort and charisma. We use virtual babies because they are easily perceived as genderless~\cite{condry1976sex} and at the same time there is literature on gender bias observed in real life that can be used as comparison.



\section{Related Work}
\label{sec:relatedWork}

According to~\cite{scott2007gender, wallach2010gender}, gender is a social and cultural construction 
built on stereotypes. For example, a pink shirt is assigned to a girl, while a blue shirt is assigned to a boy. So, a woman self-identify
as female had her construction of her feminine self based on a social standard of femininity. In this sense, in the work of~\cite{will1976maternal}, the authors carried out a perceptual experiment with two groups of mothers, one group receiving the male baby with a male name and wearing an outfit considered male, and the other group receiving the male baby with a female name and dressed in clothing considered feminine. In one of the results, the group that received the baby with a female name presented more a doll than a train. 
In the two work by~\cite{zibrek2015exploring, zibrek2013evaluating}, the authors investigated the perception of gender of virtual humans in relation to different emotional feelings, and measured the effect on people's perception. The results indicated that participants classify the gender according to the emotion showed by the character.
In this case, it was more evident when the motion had a stereotyped gender movement than when the character walked based on real human motion
. In the work of~\cite{bailey2017gender}, the authors assessed gender differences in the perception of avatars, and the results indicated that gender is important in the perception of emotions. Similar to the work of~\cite{nag2020gender}, where the authors conducted a study to assess people's perception of male, female, and androgynous virtual humans. The authors evaluated the stereotyped assumptions of gender traits and roles in virtual humans. Results showed that gender stereotypes in the virtual humans were not perceived, and the androgynous virtual humans were perceived as a middle ground between genders stereotypes. It is important to emphasize that this work used other information to present the gender, as hair and neck size. By the other hand, the work by~\cite{garza2019emotional} presented a methodology to perform content analysis on the representation of characters in children's 3D animation movies. The authors noted that female characters and their emotional expressions are still developed to fit into patterns of social stereotypes, which are "easier" to introduce into 3D animated children's movies.
\cite{seyama2007uncanny} investigated UV by measuring human perceptions of facial images with a high level of realism. In addition, the authors evaluated whether there was a correlation between the gender classification of virtual humans and UV, but there were no significant results. In the work by~\cite{mcdonnell2012render}, the authors evaluated whether perception towards virtual humans can be affected by rendering styles of virtual humans. One of the results showed that people had more correct answers when they saw female virtual humans than male ones.

Human perception is important for the design of virtual humans~\cite{zell2019perception}. Charisma is in another universe of perceptive features, for example,~\cite{weber1947legitimate} identifies charisma as a type of leadership, which one entity can exercise over another in terms of worship. 
This worship is a kind of human trait that can be observed in artificial beings, as shown in the work of~\cite{macdorman2005mortality} and the work of~\cite{rosenthal2015individuals}. According to~\cite{west1980charisma}, a trait that can be more simply perceived in some sort of of fiction. According to the work of~\cite{goethals2014kings}, and the work of~\cite{awamleh1999perceptions}, the appearance and posture of a charismatic entity can have some sort of emotional impact on the viewer. 
In work of~\cite{groves2005gender}, the results showed that women are considered more charismatic than men. In~\cite{araujo2021perceived, araujo2021analysis}, the results showed that CG characters are considered more charismatic in videos (animations) than in still images. The results also showed that, for both women and men, adult female characters can be considered more charismatic than male characters. In addition, men perceived more charisma in unrealistic than realistic characters. 

\section{Proposed Methodology}
\label{sec:gender_bias_methodology}

The proposed methodology is presented in two sections, firstly in Section~\ref{sec:gender_bias_stimuli}, we present the creation of stimuli used in this work, and in the Section~\ref{sec:gender_bias_questionnaire} we detail the applied questionnaire.

\subsection{Creation of Stimuli}
\label{sec:gender_bias_stimuli}

First, explaining Condry and Condry's experiment, the authors presented to participants a video of a real 9-month-old baby sitting in a baby chair facing a mirror (with a camera mounted behind the mirror) reacting to four stimuli (the teddy bear, a jack-in-the-box, a doll, and a buzzer). The baby had neutral clothes and no accessories to avoid gender stereotypes. Then, the authors separated the participants into two groups, firstly a group receiving the baby with a female name, and the other group receiving the baby with a male name. Thus, the authors asked the participants to use predefined scales to assess the baby's emotional and semantic levels in relation to the four stimuli presented. The main goals of the authors were to know whether the baby's gender 
influenced the perception of emotional and semantic levels of the participants. 

\begin{figure*}[ht]
  \centering
  \subfigure[fig:baby_environment][Ball]{\includegraphics[width=0.32\textwidth]{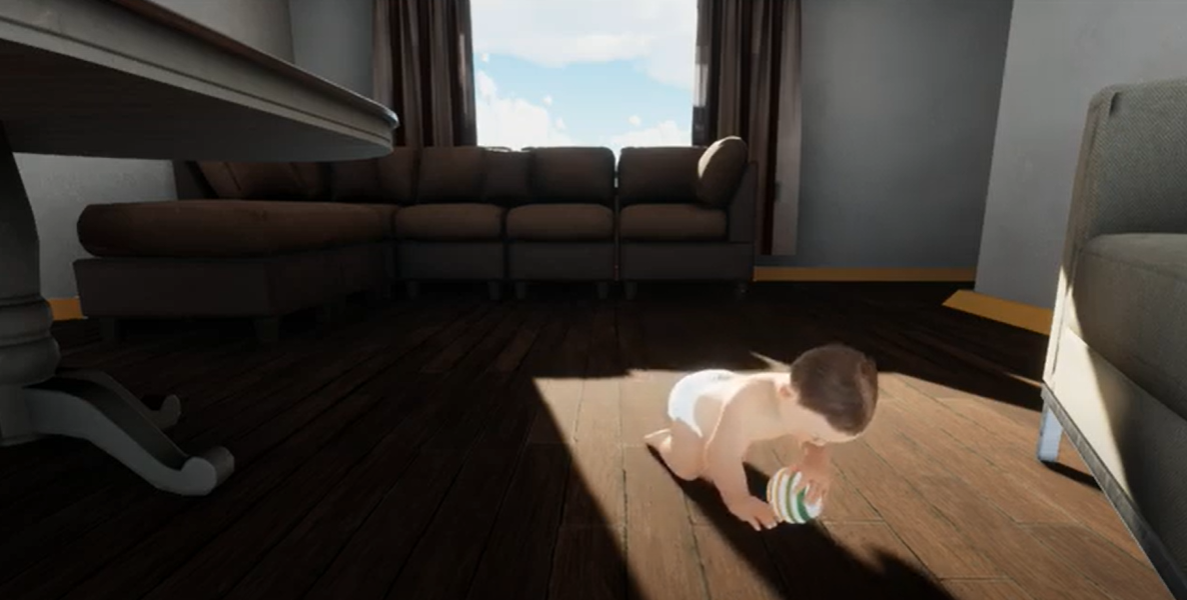}}
  \subfigure[fig:baby_environment][Unicorn]{\includegraphics[width=0.32\textwidth]{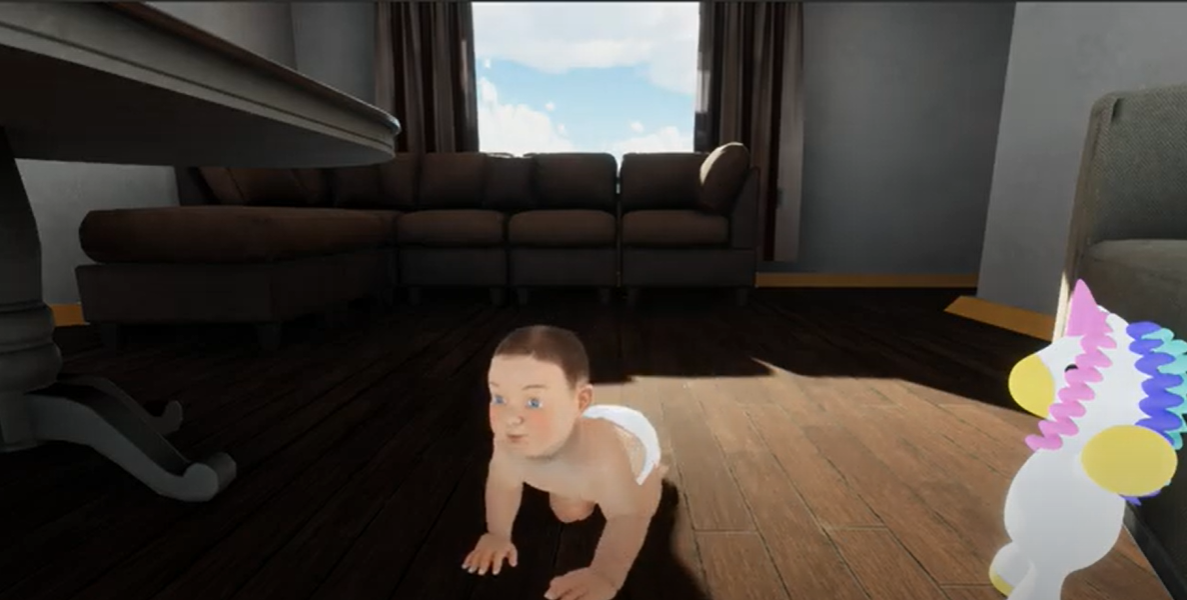}}
  \subfigure[fig:baby_environment][Jack-in-the-box]{\includegraphics[width=0.32\textwidth]{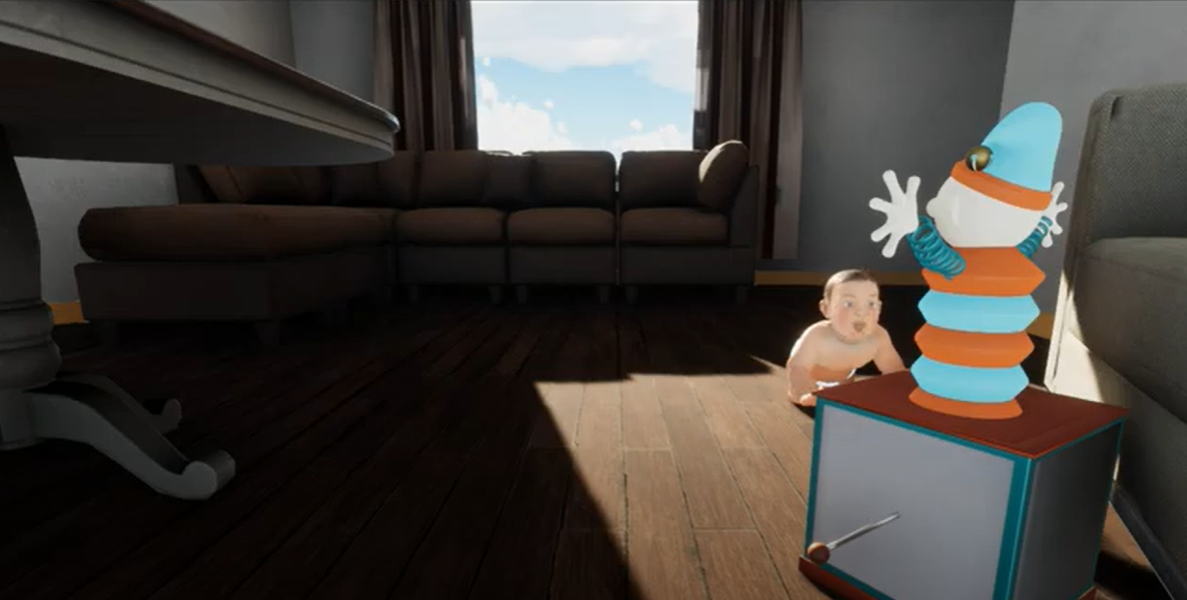}}
    \caption{Environment and three objects that the baby interacts with: (a) the baby plays with the ball; (b) the baby crawls in a direction opposite to the unicorn; (c) the baby has a negative emotion seeing Jack jump out of the box (Jack-in-the-box).}
    \label{fig:baby_environment}
\end{figure*}

In the present work, we used a 3D model of a baby purchased on website\footnote{https://www.cgtrader.com/3d-models/character/child/game-ready-baby} to replicate the original work
. The model has animations of crawling, walking and playing with a ball. 
In the original work
, the authors reported that the baby reacted positively (smiled, laughed, reached out) to the teddy bear and the doll, and reacted negatively (turned away, stared, cried) to the other two objects. As facial 
reactions 
can evoke strangeness in some participants, 
our decision was to avoid explicitly facial animations but focusing on movement and animations.  Our goal here is to avoid all interference that could affect the human perception. 
Three stimulus objects were used for the virtual baby to interact. 
Firstly, the animation of the baby playing with a ball, which in our hypothesis was perceived as a positive reaction from the baby. 
Secondly, the negative reaction was hypothetically created using the same object than in the original work, i.e., a virtual model of a jack-in-the-box
\footnote{https://www.blendswap.com/blend/27680}, which contains an animation of "Jack" jumping out of the box and back into the box. The crawl animation was used for the virtual baby to reach the jack-in-the-box, and a simple facial animation (mouths opened as in a surprise) 
was created (using the facial blendshapes of the baby). 
Regarding the object that caused a hypothetically neutral reaction, a 3D model of a colored unicorn\footnote{https://free3d.com/3d-model/unicorn-doll-772526.html} was used, and the virtual baby's reaction was to crawl in the opposite direction to the unicorn, that is, without interest in the object. 
The objects used in this work can be seen in Figure~\ref{fig:baby_environment}.

Three videos as stimuli were created, each video for each object
, to present to the participants. For all videos, a virtual scenario was created to put context in the scene. 
The set contains a room which has a table, a sofa, an armchair, a curtain, an open window showing the sky with clouds, and a wooden floor (all of these objects in the set were obtained from the internet). The scenario can be seen in
Figure~\ref{fig:baby_environment}. The camera always remains in the same position pointed at the window. At the beginning of the videos, the virtual baby always starts facing the camera and objects, and with his back to the window. Also, the baby never leaves the camera view. 
The videos duration are
between $6$ and $19$ seconds, and were sent to YouTube to be added to the questionnaire. 

\subsection{Questionnaire}
\label{sec:gender_bias_questionnaire}

The questionnaire was also based on the work of Condry and Condry
. First, the participants were presented with the consent form approved by the Ethics Committee of our University\footnote{Information about the Ethics committee and the project name were omitted for blind review}. 
Therefore, participants were informed of potential risks. 
In addition, the questionnaire was created on the Qualtrics platform\footnote{{https://www.qualtrics.com}}, 
distributed on social networks, and 
the participants were volunteers. Participants were also asked about their demographics information: age, educational level, gender, and familiarity with CG (games, movies, simulations, etc). 
After this, three types of questionnaire were created, one containing the baby with a female name, another containing the baby with a male name, and finally one containing the unnamed baby. When one participant used the link to answer the test, we randomly select one of the questionnaires. 
Both 
names were presented in an introductory text block, which presented the baby with the name and gender (no name and gender in the unnamed group), and stating that the baby was 9 months old. While in the work of Condry and Condry there were two groups, 
we decided to also test the possibility of unnamed baby, 
in order 
to assess whether people perceive that an unnamed baby does not have a defined gender. 


As in the original study, this present work also used emotional rating scales. 
For each video
, the participants were instructed to rate, on 11-Likert Scales, the pleasure, anger and fear felt by the baby
. If the participant perceived that an emotion did not appear in the video, then she/he was instructed to score the lowest value on the scale (with 0 = "Absence of Emotion"), and the highest value if the emotion appeared as much as possible (with 10 = "High Intensity"). This step aims to test the $H0_{1}$ hypothesis ("defining that women and men perceive emotions similarly in female, male and unnamed animated virtual babies"), and it provides a 2 (Women and Men responses) x 3 (Ball, Jack-in-the-box and Unicorn) x 3 (Pleasure, Anger, Fear) x 3 (Female name, Male name, and Unnamed) design structure.

To investigate the $H0_{3}$ ("defining that women and men perceive comfort and charisma similarly in baby with male name and unnamed baby, and different in relation to baby with female name"), we followed the same question structure of the emotional scales using 11-Likert scales, but containing only two questions (we called them perceptual scales), one about perceived comfort and the other about perceived charisma. The comfort question was "How comfortable are you with the virtual baby? For example, did you feel any discomfort (strangeness) while watching the video? (0 = Totally Uncomfortable, and 10 = Totally Comfortable)". 
Regarding the question of perceived charisma, 
the question given to participants was "What is the level of charisma displayed by the virtual baby? (0 = Not at all Charismatic, and 10 = Very Charismatic)".  
The comfort and charisma questions were inspired by some work of literature~\cite{mori2012uncanny, araujo2021perceived, araujo2021analysis, katsyri2015review}. This step had a 2 (Women and Men responses) x 3 (Ball, Jack-in-the-box and Unicorn) x 2 (Comfort and Charisma) x 3 (Female name, Male name, and Unnamed) design structure.

Finally, three questions were asked to the participants, and also based on the original work: \textit{i)} "Did the baby have a name? If so, what was the name?", being a question with an open text answer; \textit{ii)} "How old was the baby?", with nine 
possible answers, with "Less than 7 months old" being the first one, "More than 12 months old" being the penultimate one, and "I don't know how to inform" the last one; and finally \textit{iii)} "What was the baby's gender?", with "Female", "Male", and "I don't know" as possible answers. However, only the question of item \textit{iii)} was part of our focus in this work, as it aimed to test hypothesis $H0_{2}$ ("defining that the unnamed baby is recognized as genderless"). 
This step had a 2 (Women and Men responses) x 3 (Ball, Jack-in-the-box and Unicorn) x 3 (questions about name, age, and gender) x 3 (Female name, Male name, and Unnamed) design structure.

\section{Results}
\label{sec:gender_bias_results}

This section presents the analysis and results of the questionnaire responses presented in Section~\ref{sec:gender_bias_methodology}. In total, the questionnaire was answered by 168 volunteers through social networks, but only 148 responded completely, being 79 women, 66 men and three people who chose another option\footnote{We removed these three participants because this groups was very small.} 
Regarding age, from 145, 105 people were younger than 36 years old. 
Regarding education, 
105 people completed undergraduate studies. 
In addition, 130 responded that they were familiar with CG. Regarding the statistical analyses, we used the tests: \textit{Kruskal-Wallis} (in this case we used the \textit{Bonferroni} correction as a post hoc test), \textit{Mann-Whitney}, and \textit{Chi-Square}. These methods were used through the Statsmodels\footnote{{https://www.statsmodels.org/dev/index.html}} and SciPy\footnote{{https://docs.scipy.org/doc/scipy/index.html}} libraries of the Python language.
In all analysis, we used a significance level of $.05$. This section is divided as follows: \textit{i)} Section~\ref{sec:analysis_emotional_perception}, presents the results regarding the responses of the emotional scales presented in the questionnaire; \textit{ii)} Section~\ref{sec:analysis_gender_categorization} discusses the participants' answers regarding the perceived virtual human gender; 
and finally \textit{iii)} Section~\ref{sec:analysis_comfort_charisma} presents results about the perceived comfort and charisma.

\subsection{Analysis of Emotional Perception}
\label{sec:analysis_emotional_perception}

Explaining the results related to hypothesis $H0_{1}$, Table~\ref{tab:gender_bias_survey} presents the averages of the emotional rating scales, where the lines present the values referring to the separate videos (Ball, Unicorn, and Jack-in-the-box) and 
to all of them together (All = Ball+Unicorn+Jack-in-the-box). The columns represent the separate emotional scales (Pleasure, Anger, and Fear), and all together (Emotions = Pleasure+Anger+Fear). The first four data columns refer to the answers of the participants who received the virtual baby that had the female name, followed by the 
columns with the results regarding the answers of the participants who received the virtual baby that had the male name, and followed by the 
columns referring to to the unnamed baby. In addition, in the last columns 
there are the results of the statistical analysis performed using the Kruskal-Wallis test (non-parametric), which measures the comparison between the responses of the three groups (values referring to the female name x values referring to the male name x values referring to the to the unnamed). Therefore, 
the first lines refer to women answers, the middle five lines refer to men answers, and the last lines refer to \textit{p}-values of \textit{Mann-Whitney} test of comparisons between data of woman and men. Remembering that the mean values can vary between 0 and 10, which were the values of the 11-Likert Scales used in the questions.



\begin{table*}[ht]
  \centering
  \caption{Table of results referring to the averages of emotional rating scales. The first four data columns correspond to the responses of the group of participants that received the questionnaire containing the baby with a female name, the next four columns correspond to the responses of the group that received the questionnaire containing the baby with a male name, and the next four columns were referring to the answers on the questionnaire that had the unnamed baby. In addition to ratings on the three videos (Ball, Unicorn, Jack* = Jack-in-the-box), All* is about rating the three videos together, or about rating against all emotions together (Pleasure+Anger+Fear). Regarding the last three columns, the \textit{p}-values referring to the \textit{Kruskal-Wallis} test are presented, comparing the response values of the three groups, that is, the group that received the baby with a female name versus the group that received the baby with male name versus group that received the unnamed baby. In addition, the first five lines refer to women's data,  the middle five lines refer to men's data, and the last five lines present the \textit{p}-values referring to the \textit{Mann-Whitney} test of the comparisons between the values of the lines of the women's data versus the values of the lines of the men's data. In bold we highlight the \textit{p}-values smaller or equal than $.05$. }
  \label{tab:gender_bias_survey}
  \begin{adjustbox}{max width=0.99\linewidth}
  \begin{tabular}{|c|c|c|c|c|c|c|c|c|c|c|c|c|c|c|c|c|c|c|}
   \hline
    \multicolumn{19}{|c|}{\textbf{Women}}\\
    \hline
     & \multicolumn{4}{c|}{\textbf{Female Name}}  & & \multicolumn{4}{c|}{\textbf{Male Name}}& & \multicolumn{4}{c|}{\textbf{Unnamed}}&&\multicolumn{3}{c|}{\textbf{\textit{Kruskal-Wallis} test}} \\
    \hline
     \textbf{Stimuli} & \textbf{Pleasure} & \textbf{Anger} & \textbf{Fear} & \textbf{All*} &  &\textbf{Pleasure} & \textbf{Anger} & \textbf{Fear} & \textbf{All*}&&\textbf{Pleasure} & \textbf{Anger} & \textbf{Fear} & \textbf{All*}&&\textbf{Emotion} & \textbf{Stats} & \textbf{\textit{p}-value} \\
     \hline
    \textbf{Ball} & $6.96$ & $1.72$ & $1.24$ & - & - & $5.66$ & $1.04$ & $0.91$ & - & - & $4.80$ & $0.90$ & $0.50$ & - & - & \textbf{Pleasure} & $10.19$ & $\mathbf{.006}$ \\ 
     \hline
     \textbf{Unicorn} & $2.64$  & $0.72$ & $1.24$ & - & - & $3.16$ & $0.41$ & $1.12$ & - & - & $2.33$ & $0.63$ & $1.16$ & - & - & \textbf{Anger} & $1.91$ & $.38$\\ 
    \hline
     \textbf{Jack*} & $5.96$ & $0.52$ & $2.48$ & - & - & $4.37$ & $0.54$ & $2.12$ & - & - & $3.40$ & $1.00$& $2.66$ & - & - & \textbf{Fear} & $0.77$ & $.67$\\
    \hline
     \textbf{All*} & $5.18$ & $0.98$ & $1.65$ & $2.60$ & - & $4.40$ & $0.66$ & $1.38$ & $2.15$ & - & $3.51$ & $0.84$& $1.44$ & $1.93$ & - & \textbf{All*} & $7.00$ & $\mathbf{.03}$ \\ 
    \hline
    \multicolumn{19}{|c|}{\textbf{Men}}\\
    \hline
     & \multicolumn{4}{c|}{\textbf{Female Name}}  & & \multicolumn{4}{c|}{\textbf{Male Name}}& & \multicolumn{4}{c|}{\textbf{Unnamed}}&&\multicolumn{3}{c|}{\textbf{\textit{Kruskal-Wallis} test}} \\
    \hline
     \textbf{Stimuli} & \textbf{Pleasure} & \textbf{Anger} & \textbf{Fear} & \textbf{All*} &  &\textbf{Pleasure} & \textbf{Anger} & \textbf{Fear} & \textbf{All*}&&\textbf{Pleasure} & \textbf{Anger} & \textbf{Fear} & \textbf{All*}&&\textbf{Emotion} & \textbf{Stats} & \textbf{\textit{p}-value} \\
     \hline
    \textbf{Ball} & $6.08$ & $0.56$ & $0.60$ & - & - & $6.00$ & $1.95$ & $0.86$ & - & - & $6.55$ & $0.75$ & $3.05$ & - & - & \textbf{Pleasure} & $1.88$ & $.38$ \\ 
     \hline
     \textbf{Unicorn} & $2.64$  & $0.40$ & $0.40$ & - & - & $3.26$ & $0.56$ & $1.17$ & - & - & $2.70$ & $0.75$ & $1.95$ & - & - & \textbf{Anger} & $3.54$ & $.16$\\ 
    \hline
     \textbf{Jack*} & $4.32$ & $0.68$ & $1.96$ & - & - & $5.56$ & $0.60$ & $3.52$ & - & - & $6.10$ & $1.04$& $0.90$ & - & - & \textbf{Fear} & $6.57$ & $\mathbf{.01}$\\
    \hline
     \textbf{All*} & $4.34$ & $0.54$ & $0.98$ & $1.96$ & - & $4.94$ & $1.04$ & $1.85$ & $2.61$ & - & $5.11$ & $0.88$& $2.03$ & $2.67$ & - & \textbf{All*} & $7.96$ & $\mathbf{.01}$ \\
    \hline
    \multicolumn{19}{|c|}{\textbf{Men vs. Women - \textit{Mann Whitney} test}}\\
    \hline
     & \multicolumn{4}{c|}{\textbf{Female Name}}  & & \multicolumn{4}{c|}{\textbf{Male Name}}& & \multicolumn{4}{c|}{\textbf{Unnamed}}&&\multicolumn{3}{c|}{\textbf{}} \\
    \hline
     \multirow{2}{*}{\textbf{Stimuli}} & \textbf{Pleasure} & \textbf{Anger} & \textbf{Fear} & \textbf{All*} &  &\textbf{Pleasure} & \textbf{Anger} & \textbf{Fear} & \textbf{All*}&&\textbf{Pleasure} & \textbf{Anger} & \textbf{Fear} & \textbf{All*}&& &  &  \\

     & \textbf{\textit{p}-value} & \textbf{\textit{p}-value} & \textbf{\textit{p}-value} & \textbf{\textit{p}-value} &  &\textbf{\textit{p}-value} & \textbf{\textit{p}-value} & \textbf{\textit{p}-value} & \textbf{\textit{p}-value}&&\textbf{\textit{p}-value} & \textbf{\textit{p}-value} & \textbf{\textit{p}-value} & \textbf{\textit{p}-value}&&&& \\
     \hline
    \textbf{Ball} & $.31$ & $\mathbf{.008}$ & $.21$ & - & - & $.36$ & $.25$ & $.29$ & - & - & $\mathbf{.02}$ & $.33$ & $0.15$ & - & - & - & - & - \\ 
     \hline
     \textbf{Unicorn} & $.31$  & $\mathbf{.05}$ & $\mathbf{.01}$ & - & - & $.34$ & $.33$ & $.14$ & - & - & $.21$ & $.41$ & $.11$ & - & - & - & - & -\\ 
    \hline
     \textbf{Jack*} & $\mathbf{.04}$ & $.33$ & $.45$ & - & - & $.14$ & $.35$ & $\mathbf{.04}$ & - & - & $\mathbf{.005}$ & $.42$& $.45$ & - & - &  - & - & -\\
    \hline
     \textbf{All*} & $.07$ & $\mathbf{.01}$ & $.08$ & $\mathbf{.006}$ & - & $.17$ & $.20$ & $\mathbf{.03}$ & $\mathbf{.01}$ & - & $\mathbf{.002}$ & $.48$& $.12$ & $\mathbf{.01}$ & - &  - & - & -\\
    \hline
  \end{tabular}
  \end{adjustbox}
\end{table*}

\textbf{With regard to women population}, the obtained data in  Table~\ref{tab:gender_bias_survey} showed that, in general, emotional perception was higher for the baby with a female name than for the other groups in mostly cases.
In particular, applying the statistical \textit{Kruskal-Wallis} test with the global average values, 
we can see that there were significantly different values in pleasure and in global analysis (highlighted in bold). Furthermore, using a \textit{Bonferroni correction} as a \textit{Post Hoc} test, in both cases, 
the women's perception of emotion ($.03$) and pleasure ($.004$) had the greatest difference in the comparison between the group that received the female name and the group that received the unnamed baby. 
Still regarding the answers from women, when we analysed the stimulus videos separately\footnote{This data are not in Table~\ref{tab:gender_bias_survey} due to lack of space.}, we only found significant \textit{p}-values in the ball ($.02$) and Jack-in-the-box ($.03$) stimuli, in the comparisons of the three groups of perceptions about pleasure. 
Therefore, \textbf{we can say that the women perceived that the baby with female name was more emotional, in general, and felt more pleasure than the baby with male name and the unnamed baby.}

\textbf{With regard to the results of men}, the opposite happened
, that is, in most cases the mean values were lower for participants who received the virtual baby with a female name. In the statistical comparison, the \textit{p}-values were significant in fear
, and in relation to emotions in general. Regarding the \textit{Post Hoc} test, both in the perception of all emotions ($.01$) and in the perception of fear ($.03$), the most different groups were the groups of female and male names. 
Analyzing only the videos separately, we only found a significant result ($.02$) in the perception of fear in the unicorn stimulus, and the biggest perceptual difference was between the group that received the female name and the group that received the unnamed baby ($.03$). Therefore, \textbf{we can say that the men perceived both the baby with male name and the unnamed baby were more emotional and felt more fear than the baby with female name. In general, these cases were more evident when we compared men's perception of the baby with male and female names.} Bearing in mind that for both women and men, the baby is the same, only the name were changed.

To measure the \textbf{difference between the perception of women and men}, we performed a \textit{Mann-Whitney} test between the values. Regarding emotions in general, we found significant values in the three cases, that is, in the comparison between women and men in the three groups
. Interpreting these three values and looking at the averages of general emotions in  Table~\ref{tab:gender_bias_survey} 
we can say that \textbf{women perceived more emotion in the babies with a female name than men, and men perceived more emotion in the babies with a male name and unnamed than women}.

\subsection{Analysis of Gender Categorization in Virtual Humans}
\label{sec:analysis_gender_categorization}

Regarding $H0_{2}$ (defining that the unnamed baby is recognized as genderless), 
in all cases, both when the baby had a female name and a male name, only five women and three men answered wrong
. However, when the baby was not assigned a gender, 
18 women and 11 men responded that the baby was male. We used the \textit{Chi-Square} test to compare the correct answers between the three groups, that is, 
"female" for 
who received the female name, "male" for 
the male name, and "I don't know" for 
without name. 
Comparing women's and men's correct answers, we found no significant results, that is, \textbf{women and men were similarly correct when the babies were assigned a gender (female and male names), and were wrong in a similar way when the baby was not assigned a gender (unnamed baby).} \textbf{This is an indication that even if the baby 
does not have visual gender stereotypes, people will still point out that it is male.}

\subsection{Perceived Comfort and Charisma}
\label{sec:analysis_comfort_charisma}

Explaining the results related to hypothesis $H0_{3}$ (defining that women and men perceive comfort and charisma similarly in baby with male name and unnamed baby, and different in relation to baby with female name~\footnote{The goal was to compare with the results provided in~\cite{araujo2021analysis}.}), Table~\ref{tab:perceptual_scales} presents the averages of the perceptual scales and the statistical analysis
. The different columns state for the perceived comfort and charisma. 

\begin{table*}[ht]
  \centering
  \caption{Table of results referring to the averages of perceptual scales (Comfort and Charisma).
  Regarding the last three columns, the \textit{p}-values referring to the \textit{Kruskal-Wallis} test are presented, comparing the response values of the three groups.
  In addition, the last five lines present the \textit{p}-values referring to the \textit{Mann-Whitney} test of the comparisons between the values of the lines of the women's data versus the values of the lines of the men's data. In bold, \textit{p}-values smaller than or equal to $.05$. }
  \label{tab:perceptual_scales}
  \begin{adjustbox}{max width=0.99\linewidth}
  \begin{tabular}{|c|c|c|c|c|c|c|c|c|c|c|c|c|}
   \hline
    \multicolumn{13}{|c|}{\textbf{Women}}\\
    \hline
     & \multicolumn{2}{c|}{\textbf{Female Name}}  & & \multicolumn{2}{c|}{\textbf{Male Name}}& & \multicolumn{2}{c|}{\textbf{Unnamed}}&&\multicolumn{3}{c|}{\textbf{\textit{Kruskal-Wallis} test}} \\
    \hline
     \textbf{Stimuli} & \textbf{Comfort} & \textbf{Charisma} &  & \textbf{Comfort} & \textbf{Charisma} & & \textbf{Comfort} & \textbf{Charisma} & & \textbf{Perceived} & \textbf{Stats} & \textbf{\textit{p}-value} \\
     \hline
    \textbf{Ball} & $7.40$ & $6.60$ & - & $6.50$ & $5.87$ & - & $6.70$ & $4.76$ & - & \textbf{Comfort} & $5.06$ & $.07$ \\ 
     \hline
     \textbf{Unicorn} & $7.20$ & $6.04$ & - & $5.41$ & $4.79$ & - & $6.36$ & $4.56$ & - & \textbf{Charisma} & $13.49$ & $\mathbf{.001}$\\ 
    \hline
     \textbf{Jack*} & $7.08$ & $6.68$ & - & $5.58$ & $5.12$ & - & $6.53$ & $4.63$ & - & - & - & -\\
    \hline
     \textbf{All*} & $7.22$ & $6.44$ & - & $5.83$ & $5.26$ & - & $6.53$ & $4.65$ & - & - & - & -\\ 
    \hline
    \multicolumn{13}{|c|}{\textbf{Men}}\\
    \hline
     & \multicolumn{2}{c|}{\textbf{Female Name}}  & & \multicolumn{2}{c|}{\textbf{Male Name}}& & \multicolumn{2}{c|}{\textbf{Unnamed}}&&\multicolumn{3}{c|}{\textbf{\textit{Kruskal-Wallis} test}} \\
    \hline
     \textbf{Stimuli} & \textbf{Comfort} & \textbf{Charisma} &  & \textbf{Comfort} & \textbf{Charisma} & & \textbf{Comfort} & \textbf{Charisma} & & \textbf{Perceived} & \textbf{Stats} & \textbf{\textit{p}-value} \\
     \hline
    \textbf{Ball} & $6.92$ & $6.36$ & - & $7.43$ & $6.60$ & - & $6.85$ & $5.80$ & - & \textbf{Comfort} & $3.94$ & $.13$ \\ 
     \hline
     \textbf{Unicorn} & $6.20$ & $4.56$ & - & $7.30$ & $6.30$ & - & $6.60$ & $5.80$ & - & \textbf{Charisma} & $3.59$ & $.16$\\ 
    \hline
     \textbf{Jack*} & $5.72$ & $5.52$ & - & $6.78$ & $6.21$ & - & $6.15$ & $6.20$ & - & - & - & -\\
    \hline
     \textbf{All*} & $6.28$ & $5.48$ & - & $7.17$ & $6.37$ & - & $6.53$ & $5.93$ & - & - & - & -\\ 
    \hline
    \multicolumn{13}{|c|}{\textbf{Men vs. Women - \textit{Mann Whitney} test}}\\
    \hline
     & \multicolumn{2}{c|}{\textbf{Female Name}}  & & \multicolumn{2}{c|}{\textbf{Male Name}}& & \multicolumn{2}{c|}{\textbf{Unnamed}}&&\multicolumn{3}{c|}{\textbf{}} \\
    \hline
     \multirow{2}{*}{\textbf{Stimuli}} & \textbf{Comfort} & \textbf{Charisma} &  & \textbf{Comfort} & \textbf{Charisma} & & \textbf{Comfort} & \textbf{Charisma} & & &  &  \\

     & \textbf{\textit{p}-value} & \textbf{\textit{p}-value} &  & \textbf{\textit{p}-value} & \textbf{\textit{p}-value} & & \textbf{\textit{p}-value} & \textbf{\textit{p}-value} &&&& \\
     \hline
    \textbf{Ball} & $.23$ & $.41$ & - & $.27$ & $.20$ & - & $.46$ & $.11$ & - & - & - & - \\ 
     \hline
     \textbf{Unicorn} & $.09$ & $\mathbf{.04}$ & - & $\mathbf{.02}$ & $\mathbf{.03}$ & - & $.47$ & $\mathbf{.05}$ & - & - & - & - \\  
    \hline
     \textbf{Jack*} & $\mathbf{.02}$ & $.07$ & - & $.17$ & $.12$ & - & $.24$  & $\mathbf{.01}$ & - & - & - & - \\ 
    \hline
     \textbf{All*} & $\mathbf{.009}$ & $\mathbf{.02}$ & - & $\mathbf{.02}$ & $\mathbf{.01}$ & - & $.36$  & $\mathbf{.001}$ & - & - & - & - \\ 
    \hline
  \end{tabular}
  \end{adjustbox}
\end{table*}

Looking at the averages of women's responses
, with respect to perceived comfort, women felt more comfortable when observing a baby with female name than when observing a baby with male name or an unnamed baby. Overall among the three groups, perceived comfort values were lower when women observed the baby with male name. However, we did not find significant results, that is, \textbf{there was no gender effect in relation to the comfort perceived by women.} Regarding perceived charisma, women rated the baby with female name as more charismatic than the baby with male name and the unnamed baby. In this case, the unnamed baby was always rated as the least charismatic. Overall, we found a significant difference (gender effect), being greater in the evaluative comparison between the female name and the unnamed \textit{p}-value = $.0008$).
Therefore, \textbf{in general, we can say that women rated the baby with female name as more charismatic, especially when compared to the unnamed baby.}

Regarding the comfort perceived by men, in almost all cases, men felt more comfortable with the baby with male name, while the female had the lowest comfort values
. However, we also did not find significant results, that is, \textbf{there was also no gender effect in relation to perceived comfort by men.} Regarding perceived charisma, men also rated the male as the most charismatic, and the female having most of the lowest values. Overall, we also did not find a significant result. Therefore, we can say that, \textbf{in general, as well as in terms of perceived comfort, there was no gender effect in relation to perceived charisma.}

 Comparing women and men, regarding the group that received the female name, we can notice that in all cases the average values of comfort and charisma were higher for women than for men. On the other hand, regarding the male name, in both perceived comfort and charisma, the mean values were higher in the perception of men than in the perception of women. Statistically, this differences were significant.
With respect to the unnamed baby group, the perceived comfort values were mostly higher for men than for women, but the values were very close, and therefore the results were not significantly different. While for perceived charisma, all mean values were higher for men than for women. 
\textbf{Therefore, comparing women and men,
we can say that women felt more comfortable and perceived more charisma with a baby with female name, while men had the same result in relation to the male-named baby.}

\section{Discussion}
\label{sec:discussion}

This section aims to discuss the results presented in Section~\ref{sec:gender_bias_results}.
The Hypothesis $H0_{1}$ ("defining that women and men perceive emotions similarly in female, male and unnamed animated virtual babies") was refuted in our study.
In fact, our results indicate that the baby's gender thorough a gendered name can impact the way people perceive emotions, even if that baby does not have behavioral gender stereotypes. Furthermore, our results are in agreement with the work of~\cite{condry1976sex}, who studied gender bias in real life. It is interesting to note that we apparently also maintain our bias in virtual environments.
This can be related to gender identification~\cite{scott2007gender}, for example, women identify more with female virtual humans, but this could mean the industry does not need to worry about creating stereotyped animated virtual humans to convey gender.
Our research indicates that by just informing the baby's name was enough to create a gender perception that impact the participant emotional answer. This is an indication that it is possible to deconstruct gender stereotypes through virtual humans, as mentioned by~\cite{draude2011intermediaries}. Certainly, more research with adult virtual humans are needed, however genderless aspects should be provided and verified in order to really present a genderless character (avoiding signs like hair, size of neck, mouth color, etc).

In relation to the $H0_{2}$ hypothesis ("defining that the unnamed baby is recognized as genderless"), most participants who received the unnamed baby defined that the virtual baby was male. These results are also in accordance with the work of Condry and Condry, as the unnamed baby was assessed as male, that is, there is a gender bias, both in the real life experiment and in the virtual experiment.

Regarding hypothesis $H0_{3}$ ("defining that women and men perceive comfort and charisma similarly in baby with male name and unnamed baby, and different in relation to baby with female name"), our results were compared with the work by~\cite{araujo2021analysis}. It is important to highlight that the work of Araujo analyzed the perceived comfort and charisma of women and men over characters from different media (mostly adults) created using CG. However, even with this difference, our results were similar to those of Araujo, that is, women felt more comfortable with a realistic female virtual human than a male virtual human, and men felt more comfortable with a realistic male virtual human than a female virtual human. Therefore, as in our results, Araujo's work did not find a significant result in animations, but the authors did find a significant results in terms of static images, comparable to what has been in the present work. Regarding perceived charisma, in our work, the effect was similar to perceived comfort, that is, women rated the female virtual baby as the most charismatic, and men rated the male virtual baby as the most charismatic. This was more evident for women, as the result was significant. In the work of Araujo and his colleagues, for both women and men, realistic female characters were considered more charismatic than realistic male characters. In addition, the results are in accordance also with the work from~\cite{groves2005gender} who studies the perception of charisma in real life and observed that women can be considered more charismatic than men. 
Our work shows that as in "real life", gender identification seems to impact the users perception.

In the case of~\cite{nag2020gender} work, men and women participants had similar successes and errors in relation to the three characters tested: female, male and androgynous, but most part of guesses were correct, once the results showed that the designed characters were indeed perceived consistently with
their desired appearance. In our case, mostly participants (men and women) were more wrong than correct when trying to guess the gender of the unnamed baby, in relation to the other groups.  
While we only informed (textually) the gender of a virtual human without visual identifications, that is, our virtual human was always the same, in~\cite{nag2020gender}, the authors used visual identifications to differentiate the virtual humans. It shows that visual cues impact the participants answers creating genders determination, as expected. On the other hand, the errors obtained in our work to guess the gender of unnamed baby seem consistent with a gender bias in the participants answers. 

Therefore, those who develop virtual humans for the public (from small to large industries) may consider that they do not necessarily need stereotyped animations to represent gender. For example, knowing the likely human gender bias, the CG community can revisit the stereotypical visual aspects and propose different ways of modeling and animating virtual humans, perhaps deconstructing the binary gender, or reinforcing it if desired. Remembering that the most important thing is that people of any gender identify with and have good experiences with virtual humans.

\section{Final Considerations}
\label{sec:finalConsiderations}

This work evaluated whether the perception of women and men is influenced by virtual humans assigned gender. We raised four questions:
\textit{i)} "How do people perceive genderless characters?", \textit{ii)} "Does this change with respect to participant gender?" and \textit{iii)} "Do we have a bias to perceptive gender?". Firstly, to answer these three questions, we raised hypotheses $H0_{1}$ and $H0_{2}$, and recreated the work of~\cite{condry1976sex} using a virtual baby. In our results, women perceived more emotions in a baby named as female, and men perceived more emotions in babies with a male name and unnamed babies (refuting $H0_{1}$). In this case, the baby was always the same, that is, just changing the name and the textually defined gender. Furthermore, the group of participants who received the unnamed baby defined that this baby was male, again, even though the baby did not have gender stereotypes (refuting also $H0_{2}$).
We also studied the question \textit{iv)} "Can gender categorization influence the perception of comfort and charisma of an animated virtual human?".
Our results indicate that women felt more comfortable and perceived more charisma in the virtual baby with a female name, and men in relation to the baby with a male name. In addition, the man rated the unnamed baby as more charismatic in comparison to women. These findings refute $H0_{3}$.

From our point of view, this article simultaneously brings contributions to the field of psychology and computing. For the field of psychology, the article advances showing how the process of gender attribution, already classically studied in real humans, also happens through subtle clues (such as a name) in virtual humans. Therefore, we have increased the level of evidence for this type of perceptual bias. In the field of computer sciences we show how it is not necessary to spend resources for the animation of the genre from a behavioral point of view if the genre can be suggested by other 'cheaper' features such as a mere name.

This work has some limitations. Firstly, we only present three animations hypothesizing that the videos did not present any gender indication. Another hypothesis is that the virtual baby is genderless. Both of these hypothesis are based on literature~\cite{condry1976sex}, which we extended for virtual humans. 
We also did not test the human perception with respect to interactions with the virtual babies, but only videos. A final word regarding our results indicate that human perception about genderless virtual humans, even if it is a baby, already shows human bias in gender detection and attribution, just as in real life. It may impact some decisions taken by CG community for instance re-visiting the stereotyped constructions to model and animate virtual humans using visual attributes such as colors, cloth, animations, hair used to define a specific gender, or not. In future work, we intend to include other features to study perception in relation to virtual babies, but also with adults virtual humans.


\section{ACKNOWLEDGMENT}

The authors would like to thank CNPq and CAPES for partially funding this work.

\bibliographystyle{IEEEtran}
\bibliography{main}

\end{document}